\documentstyle[aps,preprint]{revtex}
\draft

\begin{document}
\title{First-principles molecular dynamics 
of liquid alkali metals\\
based on the quantal hyper-netted chain theory}
\author{Shaw Kambayashi}
\address{ Center for promotion of computational science and engineering,\\ 
Japan Atomic Energy Research Institute, 
Tokai, Ibaraki 319-11, Japan }

\author{Junzo Chihara\footnote{Author to whom correspondence should be 
addressed.\\ E-mail: chihara@c3004.tokai.jaeri.go.jp}}
\address{ Solid State Physics Laboratory, 
Japan Atomic Energy Research Institute, \\
Tokai, Ibaraki 319-11, Japan }
\date{August 5, 1995}
\maketitle

\begin{abstract}
A first-principles molecular dynamics (MD) scheme is presented on the 
basis of the density-functional (DF) theory with use of the the quantal 
hyper-netted 
chain (QHNC) approximation. The DF theory brings about exact expressions 
for the ion-electron and ion-ion radial distribution functions (RDF's) of 
an electron-ion mixture as a model of a simple liquid metal. 
These exact expressions prove that an ion-electron mixture can be 
treated as a one-component liquid interacting only via a {\it pairwise} 
interaction in the evaluation of the ion-ion RDF, and provide 
a set of integral equations: 
one is an exact integral equation for the ion-ion RDF and another 
for an effective ion-ion interaction, which depends on the ion configuration 
specified by the ion-ion RDF. Hence, after some approximations are 
introduced, the MD simulation can be performed to get the ion-ion RDF 
using the ion-ion interaction determined so as to be consistent to 
the ion-ion RDF: the MD simulation and the procedure to determine 
the effective interaction from the QHNC equation are performed iteratively. 
This MD simulation coupled with the QHNC equation (QHNC-MD method) 
for the effective interaction provides a first-principles calculation 
of structures of simple liquid metal: the ion-ion and electron-ion RDF's, 
the charge distributions of an ion and a pseudoatom, the effective 
ion-ion interaction and the ion-ion bridge function are evaluated 
in a self-consistent manner from the atomic number as the only input.

We have applied this QHNC-MD method to Li, Na, K, Rb and Cs near 
the melting temperature using upto 16000 particles for the MD simulation. 
It is found that the convergence of the effective ion-ion interaction 
is fast enough for practical application to alkali metals; two MD runs are 
enough for convergence within accuracy of 3 to 4 digits, 
if the initial effective potential is properly set up.
The structure factors, thus obtained, show excellent agreements 
with experimental data observed by X-ray and/or neutron scattering. 
\end{abstract}
\narrowtext

\section{INTRODUCTION}
The liquid alkali metals have been studied extensively in both experimental 
and theoretical sides. They can be easily used to test a theoretical 
approach as the first step, since they constitute ``simple" metals and 
``simple" liquids: furthermore, there exist many reliable experimental 
results to be compared. In the standard theory, a liquid metal is treated 
as a one-component liquid interacting via a binary effective potential, 
which is determined by the pseudopotential formalism; a pseudopotential 
is introduced either by first-principles calculations or by adjusting 
parameters involved in model potentials to some experimental results.
In this treatment, the ionic structures are determined independently of 
the electronic structures in a liquid metal. 

It is only recent that a 
liquid metal is thought of as an electron-ion mixture and the ionic 
structures are determined in a coupled manner with 
the electronic structures. One of such approaches is the 
Car-Parrinello molecular-dynamics (CP-MD) technique\cite{CPMD}, 
where a liquid metal is taken as a binary mixture of ions 
and electrons. In the CP-MD method, the electron-ion interaction is 
described by a pseudopotential to produce pseudo-wavefunctions which 
can be accurately represented by small number of plain waves. The CP-MD
method possesses advantage to avoid the difficult task of constructing an 
effective ion-ion potential required to perform the molecular-dynamics 
simulation, and afford to provide {\it ab initio} calculations of metallic 
systems in principle. However, the most serious problem in this approach is 
that the number of particles used in the simulations cannot be taken large a
nd 
a total of time steps performed in the simulations is limited to a small size 
within the present computational resources. In the present work, we 
propose another scheme of {\it first principles} molecular-dynamics 
simulation based on 
the density-functional (DF) method applied to the ion-electron mixture; 
this simulation method can be performed on the large number of particles 
to last a large size of time steps, since this scheme reduces the electron-i
on 
problem to a usual classical MD coupled with a set of integral equations 
determining an effective ion-ion interaction: a problem to 
determine the ion-configuration structure and the ion-ion 
interaction in a self-consistent manner.

Previously, we have proposed a set of integral equations for radial 
distribution functions (RDF's) in an electron-ion mixture  on the basis 
of the DF theory in the quantal hyper-netted (QHNC) 
approximation\cite{hyd,QHNC}. In this QHNC formalism, the bare electron-ion 
interaction $v_{\rm eI}(r)$ and the ionic charge $Z_{\rm I}$ are 
determined self-consistently by regarding a liquid metal as a mixture 
of nuclei and electrons\cite{NEmodel}. Already, we have 
applied this approach to 
liquid metallic hydrogen\cite{hyd}, lithium\cite{QHLi}, 
sodium\cite{QHNa}, potassium\cite{QHK} and aluminum\cite{QHAl} obtaining 
ion-ion structure factors in excellent agreements with experiments. 
The QHNC equations are derived from exact expressions for the electron-ion 
and ion-ion RDF's in an electron-ion mixture: these exact expressions 
are only formal results derived from the DF theory. 
A molecular-dynamics scheme to treat 
an ion-electron mixture can be set up on the basis of these exact 
relations, which states that the electron-ion mixture can be regarded 
as a quasi-one component liquid interacting via a ${\it pairwise}$ 
interaction in the description of the ion-ion RDF\cite{QHNC} 
(hereafter, referred to as the QHNC-MD method). 
Since the QHNC formalism is derived on the electron-ion model 
where the bound-electrons forming an ion in a liquid metal is 
assumed to be clearly distinguished from the conduction electrons 
and the overlap of the core electrons is negligible,  the QHNC-MD 
method is only applicable to simple liquid metals: its application 
to liquid alkali metals is taken as an ideal test of the QHNC-MD method. 
Thus, the QHNC-MD method has been shown to 
yield structure factors of liquid alkali metals 
in excellent agreement with experiments as the result of a 
first-principles calculation in the present work.

In Sec.~\ref{sec:hnc}, we sketch the QHNC formulation: exact expressions for 
RDF's in an electron-ion mixture are obtained from the DF 
method\cite{BR}, and the nucleus-electron model is shown to 
provide a bare electron-ion interaction, which should be 
determined self-consistently.  The procedure to perform the MD simulation 
based on the QHNC theory is shown in Sec.~\ref{sec:md}: in the QHNC 
formulation the effective ion-ion interaction used in the MD simulation 
depends on the ionic structure specified by the ion-ion RDF.
Therefore, in the application 
of this MD scheme it is important to extrapolate the MD RDF 
beyond the truncation radius of the simulation correctly so 
as to be used in the determination of 
effective ion-ion interaction; this is exemplified by the method 
described in Sec.~\ref{sec:md}. Numerical procedure of the QHNC-MD 
method and the results of its application to alkali liquid metals 
are described in Sec.~\ref{sec:appl}.
The last section is devoted to a discussion, where 
the advantage/disadvantage 
of the QHNC-MD method against the CP-MD method is argued also.

\section{QUANTAL HYPER-NETTED CHAIN THEORY}
\label{sec:hnc}
A simple liquid metal can be thought of as a binary mixture of ions 
with a definite ionic charge $Z_{\rm I}$ and the conduction electrons; 
the interactions $v_{ij}(r)$ between particles [$i,j=$I or e] 
are taken as  pair-wise. The ions constitute a classical fluid, 
while the conduction electrons form a quantum fluid.
Let us refer to this mixture as the ion-electron model for a liquid metal.
Since the ions are regarded as classical particles in the electron-ion
model, the ion-ion and electron-ion RDF's become identical with the 
ion- and electron-density distributions around a fixed ion in the mixture,
 respectively\cite{QHNC}.  Because a fixed ion causes external potentials
acting on ions and electrons in the homogeneous mixture, 
the DF theory can give the exact expressions 
for the ion- and electron-density distributions, $n_{\rm I}(r|{\,\rm I})$
and $n_{\rm e}(r|{\,\rm I})$, 
in terms of those of noninteracting systems $n^0_i(r)$ under effective
external potentials $U_i^{\rm eff}(r)$ [$i$=I, e] 
\begin{equation}
 U_i^{\rm eff}(r)=v_{i{\rm I}}(r)+{\delta {\cal F}_{\rm int}\over\delta 
n_i(r|\,{\rm I})}-\mu_i^{\rm int} \label{eq:gvef1}\\   
\end{equation}
with the use of ${\cal F}_{\rm int}$ and $\mu_i^{\rm int}$, the 
interaction part of the intrinsic free-energy and the chemical 
potential, respectively\cite{QDCF}.
As a result, the DF theory provides exact, but formal, expressions for 
the ion-ion RDF $g_{\scriptscriptstyle\rm II}(r)$ and electron-ion 
RDF $g_{\scriptscriptstyle\rm eI}(r)$ as follows:
\begin{eqnarray}
n_0^{\rm I}g_{\scriptscriptstyle\rm II}(r) & = & n_{\rm I}(r|\,{\rm I})=
  n_{\rm I}^0(r|U_{\rm I}^{\rm eff})
  \equiv n_0^{\rm I}\exp[-\beta U_{\rm I}^{\rm eff}(r)]\;,\label{eq:gII}\\
n_0^{\rm e}g_{\scriptscriptstyle\rm eI}(r) & = & n_{\rm e}(r|\,{\rm I})=
  n_{\rm e}^0(r|U_{\rm e}^{\rm eff})
  \equiv\sum_i{|\psi_i(r)|^2\over
    \exp[\beta(\varepsilon_i-\mu_0^{\rm e})]+1}\;,\label{eq:geI}
\end{eqnarray}
where $\mu_0^{\rm e}$ denotes the chemical potential of a
non-interacting electron gas, $n_0^{\rm I}$ ($n_0^{\rm e}$) is the
number density of ions (electrons), and $\beta=(k_{\rm B}T)^{-1}$
the inverse temperature.  The electron-density distribution $n_{\rm
e}^0(r|U)$ is determined by solving the wave equation for an
electron under the external potential $U(r)$
\begin{equation}
\left[ -(\hbar^2/2m)\nabla^2+U(r) \right]\psi_i(r)
  =\varepsilon_i\psi_i(r)\;.
\end{equation}
In the similar way to the case of classical binary mixtures, the effective
external potentials $U_i^{\rm eff}(r)$ given by Eq.~(\ref{eq:gvef1}) 
are written as
\begin{eqnarray}
U_i^{\rm eff}(r) & = & 
  v_{i\rm I}(r)-{\it\Gamma}_{i\rm I}(r)/\beta
  -B_{i\rm I}(r)/\beta\;,\label{eq:Ueff}\\
{\it\Gamma}_{i\rm I}(r) & \equiv &
  \sum_l \int C_{il}(|{\bf r}-{\bf r}'|)n_0^l
  [g_{{\scriptscriptstyle l}{\rm I}}(r)-1]d{\bf r}'\;,
\end{eqnarray}
in terms of the direct correlation
functions (DCF's) $C_{ij}(r)$ and the bridge functions $B_{i\rm
I}(r)$. Here, the DCF's $C_{ij}(r)$ in the 
ion-electron mixture are defined within the framework of the DF theory
 by
\begin{equation}
C_{ij}(|{\bf r}-{\bf r'}|) \equiv -\beta { \delta^2 {\cal F}_{\rm 
int}[n_I,n_e]  \over  \delta n_{i}({\bf r})\delta n_{j}({\bf r'}) 
}\biggl |_0\;,  \label{e:dcf} 
\end{equation}
where the suffix 0 denotes the functional derivative at the uniform 
densities \cite{QDCF}.
Actually the explicit expression for the DCF's are given by the 
Fourier transform in the matrix form
\begin{equation}
\sqrt{{\cal N}}C(k)\sqrt{{\cal N}} = (\widetilde{\chi}_Q^0)^{-1} - 
(\widetilde{\chi}_Q)^{-1} \label{e:dcfq} 
\end{equation}
in terms of the density response functions, $\widetilde{\chi}_Q\equiv 
\parallel \chi_{_{ij}}(k)\parallel$ and $\widetilde{\chi}_Q^0\equiv 
\parallel 
\chi^{0i}(Q)\delta_{ij}\parallel$, of the interacting and 
noninteracting systems, respectively, with ${\cal N}\equiv \parallel 
n_0^{i}\delta_{ij}\parallel$\cite{QDCF}. 
Note here that the density-density response functions $\chi_{iI}(Q)$ 
concerning ion reduce to the  structure factors $S_{iI}(Q)$ and
 $\chi^{0I}(Q)=1$, since the ions behave as classical particles\cite{QHNC}. 
>From this definition of the DCF's the Ornstein-Zernike relations are
derived for the ion-electron mixture:
\begin{eqnarray}
g_{\scriptscriptstyle\rm II}(r) & = & 
  C_{\rm II}(r)+{\it\Gamma}_{\rm II}(r)\;,\label{eq:gIIOZ}\\
g_{\scriptscriptstyle\rm eI}(r) & = &
  \hat BC_{\rm eI}(r)+\hat B{\it\Gamma}_{\rm eI}(r)\;.\label{eq:geIOZ}
\end{eqnarray}
Here, $\hat B$ denotes an operator defined by
\begin{equation}
{\cal F}_Q[\hat B^\gamma f(r)]
  \equiv (\chi_Q^0)^\gamma{\cal F}_Q[f(r)]
  =(\chi_Q^0)^\gamma\int\exp[i{\bf Q}\cdot{\bf r}]f(r)d{\bf r}\;,
\end{equation}
for an arbitrary real number $\gamma$, and represents a
quantum-effect of the electron through the density response
function $\chi_Q^0$ of the non-interacting electron gas.  

A set of integral equations (\ref{eq:gII}) and (\ref{eq:geI}) are
exact but formal expressions, as well as all other equations 
in the above. However, the ionic charge $Z_{\rm I}$ and
the electron-ion interaction $v_{\rm eI}(r)$ must be given 
beforehand, when we apply these formula to a 
liquid metal as an ion-electron mixture.
In order to determine these quantities from first principle, a liquid metal
must be treated more fundamentally as a mixture of nuclei and electrons 
(the nucleus-electron model), where all interactions between particles are
known as pure Coulombic. In this model, input data 
in dealing with a liquid metal 
is only the atomic number $Z_{\rm A}$ to specify the material. 
For this purpose, let us consider a liquid metal as a mixture of 
$N_{\rm I}$ nuclei and $Z_{\rm A}N_{\rm I}$ electrons, 
and solve the problem to determine
 the electron-density distribution around a nucleus fixed at the origin in
this mixture. Since a fixed nucleus causes an external potential 
$U(r)=-Z_{\rm A}e^2/r$ for this mixture to produce an inhomogeneous system,
the DF theory can be applied to this problem. It should be noticed that 
DF theory contains some arbitrariness in the choice of a reference 
system to describe the system\cite{QDCF}. We can get a simple 
description of the nucleus-electron mixture if the reference system is 
chosen to be a mixture consisting of $N_{\rm I}-1$ noninteracting ions and
$Z_{\rm I}(N_{\rm I}-1)+Z_{\rm A}$ noninteracting electrons: here, each ion
is assumed to have $Z_{\rm B}$ bound-electrons with a charge distribution 
$\rho_{\rm b}(r)$ around it and an ionic charge $Z_{\rm I}\equiv 
Z_{\rm A}-Z_{\rm B}$. With use of this reference system, the DF theory
can provide an effective external potential $v_{\rm eN}^{\rm eff}(r)$
for electrons around the fixed nucleus. Then, the electron-density 
distribution $n_{\rm e}(r|{\rm N})$ around the fixed nucleus
is obtained by solving 
the wave equation for $v_{\rm eN}^{\rm eff}(r)$ in the sum of 
the bound- and free-electron parts:
\begin{equation}
n_{\rm e}(r|{\rm N})=n_{\rm e}^{0\rm b}(r|v_{\rm eN}^{\rm eff})
=n_{\rm e}^{\rm b}(r|{\rm N})+n_{\rm e}^{\rm f}(r|{\rm N})\;.  \label{eq:neN} 
\end{equation}
Hence, the bound-electron distribution $n_{\rm e}^{\rm b}(r|{\rm N})$ 
thus determined constitutes the definition of the ``ion" in 
the electron-ion model.
Furthermore, this bound-electron distribution $n_{\rm e}^{\rm b}(r|{\rm N})$ 
should be taken identical with 
the electron distribution $\rho_{\rm b}(r)$ of an ion in the 
reference system, since the ion formed around the central nucleus is 
necessary to be the same structure as any ion in the system.
Thus, we obtain a self-consistent condition to determine 
the distribution $\rho_{\rm b}(r)$ in the premise:
\begin{equation}
\rho_{\rm b}(r)=n_{\rm e}^{\rm b}(r|{\rm N})\equiv n_{\rm e}^{\rm b}(r)
\end{equation}
with the bound-electron number $Z_{\rm B}=\int\rho_{\rm b}(r)d{\bf r}$.
On the other hand, the free-electron part $n_{\rm e}^{\rm f}(r|{\rm N})$
in Eq.~(\ref{eq:neN}) is taken as the electron-ion RDF 
$n_0^{\rm e}g_{\scriptscriptstyle\rm eI}(r)$ of the electron-ion mixture 
with the free-electron density 
$n_0^{\rm e}=Z_{\rm I}n_0^{\rm I}$, and the nucleus-nucleus RDF becomes 
the ion-ion RDF $g_{\scriptscriptstyle\rm II}(r)$.

With use of this reference system, we can obtain a tractable expression of 
$v_{\rm eN}^{\rm eff}(r)$ for the wave equation to determine 
$n_e(r|N)$ by introducing some approximations 
to the exchange-correlation term involved in it\cite{NEmodel}:
\begin{equation}\label{eq:vefeN}
v_{\rm eN}^{\rm eff}(r)={\tilde v}_{\rm eI}(r)
 -{1\over\beta}\sum_l 
  \int C_{{\rm e}l}(|{\bf r}-{\bf r}'|)n_0^l
  [g_{{\scriptscriptstyle l}{\rm I}}(r')-1]d{\bf r}'\;,
\end{equation}
where $\mu_{\rm XC}(n)$ is the exchange-correlation
potential in the local-density approximation (LDA).
Note that this expression is equal to Eq.~(\ref{eq:Ueff}) 
without the electron-ion bridge functions $B_{\rm eI}(r)$ 
except that the bare electron-ion interaction
is explicitly given by 
\begin{equation}\label{eq:barveI}
{\tilde v}_{\rm eI}(r)\equiv -{Z_{\rm A}e^2\over r}
 +\int v_{\rm ee}(|{\bf r}-{\bf r}'|)n_{\rm e}^{\rm b}(r')d{\bf r}'
 +\mu_{\rm XC}(n_{\rm e}^{\rm b}(r)+n_0^{\rm e})
 -\mu_{\rm XC}(n_0^{\rm e})\;.
\end{equation}
In this way, the treatment of a liquid metal as a nucleus-ion mixture 
is shown to provide the ion-electron model, where the bare electron-ion 
interaction ${\tilde v}_{\rm eI}(r)$ and the ionic structure 
$\rho_{\rm b}(r)$ can be determined in a self-consistent manner.

With the help of the result from the nucleus-electron model, we can
derive a closed set of integral equations for the ion-electron mixture,
if we introduce the following approximations:
(A) the electron-ion
bridge function in Eq.~(\ref{eq:QHNCei}) is neglected: 
$B_{\rm eI}(r)\simeq 0$, (B) the electron-electron DCF $C_{\rm ee}(r)$
 is approximated \cite{QHNC} as
\begin{equation}\label{eq:Cee}
C_{\rm ee}(Q)=-\beta v_{\rm ee}(Q)\left[1-G^{\rm jell}(Q)\right]\;.
\end{equation}
using the local-field correction (LFC) $G^{\rm jell}(Q)$ of the jellium model
for an electron gas,
(C) the bare ion-ion potential $v_{\rm II}(r)$ is taken as pure
Coulombic, i.e., $v_{\rm II}(r)=(Z_{\rm I}e)^2/r$, 
and (D) the bare electron-ion potential is given by 
$v_{\rm eI}(r)={\tilde v}_{\rm eI}(r)$ of Eq.~(\ref{eq:barveI}).
We have called this set of equations the quantal hyper-netted chain 
equations because of the approximation $B_{\rm eI}(r)\simeq 0$ 
in Eq.~(\ref{eq:Ueff}).

\section{MOLECULAR DYNAMICS SIMULATION
 BASED ON QUANTAL HYPER-NETTED CHAIN THEORY}
\label{sec:md}
It is important to realize that the electron-ion model leads to 
the neutral-fluid model, where the ionic behavior of a liquid metal is 
taken to be the same as a neutral
 one-component fluid interacting via a binary effective 
interaction in treating the ion-ion RDF.
This neutral-fluid model is derived from the electron-ion model, when
an effective ion-ion potential is defined in such a
way that the RDF of a one-component fluid 
should become identical with $g_{\scriptscriptstyle\rm II}(r)$ of 
the electron-ion mixture: 
\begin{equation}\label{eq:gRone}
g(r)\equiv \exp[-\beta v_{\rm eff}(r)+{\it\Gamma}(r)+B(r)]
  = g_{\scriptscriptstyle\rm II}(r)\;,
\end{equation}
with use of the Ornstein-Zernike relation for a neutral one-component fluid
\begin{equation}
 g(r)-1=C(r)+{\it\Gamma}(r)\;.
\end{equation}
In the above, 
the DCF for the one-component fluid is defined by 
$n_0^{\rm I}C(Q)\equiv 1-S_{\rm II}(Q)^{-1}$
and ${\it\Gamma}(r)\equiv 
  \int C(|{\bf r}-{\bf r}'|)n_0^{\rm I}[g(r')-1]d{\bf r}'$.

Thus, we can write the explicit expression for the effective
ion-ion potential of a liquid metal in the neutral-fluid model as
\begin{equation}\label{eq:veff}
\beta v_{\rm eff}(Q)
  \equiv \beta v_{\rm II}(Q)
  -{|C_{\rm eI}(Q)|^2n_0^{\rm e}\chi_Q^0\over
    1-n_0^{\rm e}C_{\rm ee}(Q)\chi_Q^0}\;,
\end{equation}
by taking the bridge function $B(r)$ to be $B_{\rm II}(r)$ of the
electron-ion mixture.  Equation (\ref{eq:veff}) can be
interpreted within the scope of the standard pseudopotential theory
by regarding $C_{\rm eI}(r)$ as the pseudopotential
$w_b(Q)=-C_{\rm eI}(Q)/\beta$. In this way, the ion-electron model
is reduced exactly to the neutral-fluid model with a {\it binary}
effective interaction Eq.~(\ref{eq:veff}), in which 
the many-body forces are taken into account in the form of the 
linear response expression (\ref{eq:veff}), 
since the nonlinear effect in the electron screening is involved 
in terms of the electron-ion DCF, which plays the role of a nonlinear 
pseudopotential.

By noting the above relations (\ref{eq:gRone})\,--\,(\ref{eq:veff}),
the exact expressions (\ref{eq:gII})\,--\,(\ref{eq:geIOZ}) for the
electron-ion model can be
transformed into a set of integral equations:  one is the integral
equation for a one-component fluid with the effective ion-ion
potential $v_{\rm eff}(r)$
\begin{equation}\label{eq:QHNCii}
C(r)=\exp[-\beta v_{\rm eff}(r)
  +{\it\Gamma}(r)+B(r)]-1-{\it\Gamma}(r)\;,
\end{equation}
and the other an equation for the effective ion-ion interaction 
$v_{\rm eff}(r)$, that is expressed in the form of an integral 
equation for the electron-ion DCF $C_{\rm eI}(r)$:
\begin{equation}\label{eq:QHNCei}
\hat B C_{\rm eI}(r)
 =n_{\rm e}^0(r|v_{\rm eI}
     -{\it\Gamma}_{\rm eI}/\beta-B_{\rm eI}/\beta)/n_0^{\rm e}
     -1-\hat B {\it\Gamma}_{\rm eI}(r)\;,
\end{equation}
since the effective interaction $v_{\rm eff}(r)$ is given 
in terms of $C_{\rm eI}(r)$ by Eq.~(\ref{eq:veff}).
In contrast with the usual effective
potential in the pseudopotential theory, the effective potential 
(\ref{eq:veff}) depends on the ion configuration represented 
by the ion-ion RDF $g_{\scriptscriptstyle\rm II}(r)$ through the term: 
${\it\Gamma}_{\rm eI}(r) \equiv \sum_l 
  \int C_{el}(|{\bf r}-{\bf r}'|)n_0^l
  [g_{{\scriptscriptstyle l}{\scriptscriptstyle\rm I}}(r)-1]d{\bf r}'$ 
  in Eq.~(\ref{eq:QHNCei}).

Under the assumptions (A)\,--\,(D) mentioned before, the QHNC 
equations (\ref{eq:QHNCii}) and (\ref{eq:QHNCei}) enables us to 
perform an {\it ab initio} molecular-dynamics (MD) simulation 
which requires only the atomic number $Z_{\rm A}$ and thermodynamic 
states as input parameter, in principle.  
The first estimation for $v_{\rm eff}(r)$ can be
obtained with use of $C_{\rm eI}(r)$ evaluated by Eq.~(\ref{eq:QHNCei}) 
with an initial guess for $g_{\scriptscriptstyle\rm II}(r)$. 
Next, an integral equation (\ref{eq:QHNCii}) for a one-component fluid 
can be solved by performing the classical MD simulation for this 
$v_{\rm eff}(r)$ to produce new ion-ion RDF $g_{\scriptscriptstyle\rm II}(r)$; this 
is used again in Eq.~(\ref{eq:QHNCei}) to determine a 
new estimation for $v_{\rm eff}(r)$.
This process will be continued until convergence of the effective
ion-ion potential is achieved (we refer to this procedure
as the QHNC-MD method).  However, such a straight-forward
repetition of the MD simulation to solve the QHNC equations is not
practical in the viewpoint of the computational cost.  Since the
dependence of the effective ion-ion potential $v_{\rm eff}(r)$ 
on the ionic configuration is rather weak in a simple metal 
as we have shown in \cite{QHAl}, we can adopt an approximate theory for 
$B(r)$ in Eq.~(\ref{eq:QHNCii}) to get an initial $v_{\rm eff}(r)$ 
for the QHNC-MD method. For this
purpose, we take the variational modified HNC (VMHNC) equation
proposed by Rosenfeld
\cite{VMHNC}, in which the bridge function is approximated by
$B_{\rm PY}(r;\eta)$ of the Percus-Yevick equation for hard-spheres
of diameter $\sigma$ with the packing fraction $\eta=\pi n_0^{\rm
I}\sigma^3/6$.  In the VMHNC equation, the adjustable parameter
$\eta$ is determined by the following condition:
\begin{equation}
{1\over 2}n_0^{\rm I}\int[g(r)-g_{\rm PY}(r;\eta)]
  {\partial B_{\rm PY}(r;\eta)\over\partial\eta}d{\bf r}
  +{2\eta^2\over(1-\eta)^3}=0\;,
\end{equation}
where $g_{\rm PY}(r;\eta)$ is the RDF for the hard-sphere fluid
with the Percus-Yevick equation. Thus, in a similar way to the QHNC-MD
 method, the integral equation (\ref{eq:QHNCii}) in the VMHNC 
approximation is solved in a coupled manner with Eq.~(\ref{eq:QHNCei}) 
producing new effective ion-ion interaction [referred to as the 
QHNC-VM method]. 
Furthermore, an initial potential $v_{\rm eff}(r)$ 
to this QHNC-VM method can be obtained by
approximating $g_{\scriptscriptstyle\rm II}(r)$ in Eq.~(\ref{eq:QHNCei}) 
by the step function $\theta(r-a)$ with the ion-sphere radius $a=(4\pi
n_0^{\rm I}/3)^{-1/3}$.
When this final $v_{\rm eff}(r)$ from the QHNC-VM method is used as 
an input to the QHNC-MD method, the convergent result can be obtained 
by a few repetition of MD simulation.  
Finally our procedure to solve
the QHNC equation with the MD simulation (the QHNC-MD method) 
is summarized as the flow chart shown in Figure \ref{fig:flow}.
For an initial potential $v_{\rm eff}(r)$ given by
approximation $g_{\scriptscriptstyle\rm II}(r)=\theta(r-a)$, 
the QHNC-VM method 
in the {\em preparation phase} yields a good initial 
guess for the QHNC-MD method. 
Then the MD simulation is repeatedly performed to
achieve convergence of $v_{\rm eff}(r)$ in the {\em refinement
phase}.

There are two important points to be noticed regarding the MD simulation 
when applied to the QHNC-MD method. One is that the computer simulation 
provides the RDF $g(r)$ only within the half of the side length $L$ of 
the simulation cell. This causes an unavoidable truncation error 
in calculation of the Fourier transform ${\cal F}_Q[g(r)-1]$ to be used 
in the evaluation of Eqs. (\ref{eq:veff}) and (\ref{eq:QHNCei}).
The second point is that the MD simulation is performed inevitably on 
a truncated potential for a liquid metal whose effective 
ion-ion potential is accompanied by a long-ranged oscillatory tail: 
the computer simulation may yield different RDF's depending on the
cutoff radius $R_{\rm c}$ of the potential.
Recently we have proposed a precise procedure\cite{BR} to improve 
these two defects at the same time and to get the RDF in the whole range 
of distance for the full potential $v_{\rm eff}(r)$. This method can 
be applied even to the {\it small-size} simulation result for the
truncated potential $u_{\rm c}(r)$:
\begin{equation}
u_{\rm c}(r)\equiv  \left\{
  \begin{array}{ll}
  v_{\rm eff}(r)-v_{\rm eff}(R_{\rm c}) & \mbox{for}\ r<R_{\rm c} \\
  0 & \mbox{for}\ r\ge R_{\rm c} 
  \end{array}
  \right.\;.
\end{equation}
As the first step of this procedure, we extract the bridge function
from the raw MD RDF data.  For this purpose, we extend the the raw RDF
data of the MD simulation $g_{\rm MD}(r)$, by solving an integral
equation
\begin{equation}\label{eq:grex}
g(r)\equiv
  \left\{
  \begin{array}{ll}
  g_{\rm MD}(r) & \mbox{for}\ r<R \\
  \exp[-\beta u_{\rm c}(r)+{\it\Gamma}(r)] & \mbox{for}\ r\ge R
  \end{array}
  \right.\;,
\end{equation}
coupled with the Ornstein-Zernike relation, where $R$ is the
extrapolating distance ($R<L/2$). At this stage, in order to
obtain a reliable bridge function, it is essential to take $R$ as
short as about 3 to 4 interatomic spacings or simply as $R=R_{\rm
c}$ \cite{BR} and to discard the RDF data outside the distance $R$ so as
to reduce the statistical noise contained in the raw RDF data.
Using the extended $g(r)$, the bridge function $B_{\rm MD}(r)$ can
be extracted for distances where $g_{\rm MD}(r)\ne 0$ by
\begin{equation}\label{eq:MDBr}
B_{\rm MD}(r)\equiv
  \left\{
  \begin{array}{ll}
  \beta u_{\rm c}(r)+\ln[g_{\rm MD}(r)]-{\it\Gamma}(r) 
    & \mbox{for}\ r<R \\
  0 & \mbox{for}\ r\ge R
  \end{array}
  \right.\;.
\end{equation}
At the second step to get the RDF for the full potential
$v_{\rm eff}(r)$, we solve the integral equation (\ref{eq:QHNCii})
for the full potential with use of this $B_{\rm MD}(r)$ as 
an approximation to that of the full potential: this can be 
justified by the fact that the bridge
function is not sensitive to the long-range part of the potential
and becomes very weak for long-range distance \cite{BR}.

\section{APPLICATION TO LIQUID ALKALI METALS}
\label{sec:appl}
\subsection{Numerical procedure of QHNC-MD method}
Liquid alkali metals constitute ``simple" metals in the sense that
the bound electrons forming an ion are clearly distinguished from 
the conduction electrons and the overlap of the core electrons are 
negligible; the approximations (A)--(D) used in the QHNC-MD method 
become quit good ones for these metals. Therefore, 
we have applied the QHNC-MD method for five liquid alkali metals
(Li, Na, K, Rb, and Cs) near the melting point using the 
parameters specified in Table
\ref{tab:param}; the temperature and density have been chosen
to be compared with experimental data 
in \cite{ExpLi,ExpNaKCs,ExpNaK,ExpRbCH,ExpRb,ExpRbW}. 
Here, the temperature and density of 
alkali liquids are specified by two dimensionless parameters: 
the plasma parameter $\Gamma=\beta e^2/a$ and 
$r_{\rm s}=a/a_{\scriptscriptstyle\rm B}$ 
in units of the Bohr radius $a_{\scriptscriptstyle\rm B}$, 
with the average ion-sphere radius $a$.
 
In our application of  the QHNC-MD simulation to the alkali liquids, 
the local-field correction of the jellium model in Eq.~(\ref{eq:Cee}) 
is chosen to be that proposed by Geldart and Vosko \cite{GV}, since it has 
a simple structure and gives a good approximation. 
In Eq.~(\ref{eq:barveI}), the expression given by Gunnarsson 
and Lundqvist \cite{GL} is adopted as the LDA for the 
exchange-correlation potential $\mu_{\rm XC}(n)$.

After the preparation of initial effective potential by 
the QHNC-VM method, two iterations 
in the refinement phase of the
QHNC-MD method (Figure \ref{fig:flow}) are sufficient to obtain 
a convergent solution for alkali liquid metals; 16,000 particles have
been used in the first MD run (Run-1) and 4,000 particles for the
second MD run (Run-2).  The MD simulation has been performed over
50,000 time steps for Run-1 and 100,000 time steps for Run-2 
with the cubic periodic boundary conditions; 
the temperature of the system is kept constant by the isokinetic
constraint \cite{Hoover}.  The
equations of motion are integrated by a fifth order differential
algorithm \cite{Algorithm} with the time increment
${\it\Delta}t=0.0025n_0^{{\rm I}\;-1/3}(m_{\scriptscriptstyle\rm I}\beta)$ 
with the mass of an ion $m_{\scriptscriptstyle\rm I}$: 
the corresponding real time is shown in Table \ref{tab:param}. 
In each MD simulation of Run-1 and Run-2 for 
iterations, the effective potential is 
cut at the radius $R_{\rm c}$ located at the node of 
the Friedel oscillation of $v_{\rm eff}(r)$ as shown in 
Table \ref{tab:param}.  All MD simulations have been
carried out on a vector-parallel processor Monte-4 \cite{Monte4} at 
Japan Atomic Energy 
Research Institute.  The computational time required for 10,000
steps is about 30 to 50 hours for 16,000 particles including the
sampling of the RDF.

The integral equation (\ref{eq:grex}) has been solved by an 
iterative procedure introduced by Ng \cite{Ng} to extend the 
raw MD data $g_{\scriptscriptstyle\rm MD}(r)$ to the 
whole $r$ range: the extending distance $R$ in this 
procedure (\ref{eq:grex}) is taken to be 
$R_{\rm c}$ for the whole cases.  The number of grid points and step size
used in numerical integrations are 1024 points and
${\it\Delta}r=0.025a$, respectively.  Using $C(r)$ obtained by the
HNC equation as an initial input function, it takes about 10,000
iterations to achieve convergence.

In order to examine both numerical and computational efficiency of
the QHNC-MD method, we have tested the convergence of the RDF 
 by evaluating following consistency
measure for $g(r)$:
\begin{eqnarray}
{\it\Delta}g_i(r) & \equiv & g_i(r)-g_{i-1}(r)\;,\label{eq:diff} \\
|{\it\Delta}g_i| & \equiv &
  \left({4\pi n_0^{\rm I}\over
  3}\int_0^\infty|{\it\Delta}g_i(r)|^2r^2dr\right)^{1/2}\;.
  \label{eq:measure}
\end{eqnarray}
Here $g_i(r)$ is $g_{\scriptscriptstyle\rm II}(r)$ obtained by the
$i$-th MD simulation and $g_0(r)$ is that obtained by the final
step of the preparation phase.
Figure \ref{fig:gconv} shows the consistency
measure (\ref{eq:diff}) of the present QHNC-MD simulation for 
the case of liquid Li, yielding the values $|{\it\Delta}g_1|
=1.53\times 10^{-1}$ and $|{\it\Delta}g_2|=5.92\times 10^{-3}$.  
It is easily seen that the
convergence of $g_{\scriptscriptstyle\rm II}(r)$ is very fast;
the difference of $g_{\scriptscriptstyle\rm II}(r)$ 
between Run-1 and Run-2 is
situated almost within the statistical error of the sampling of the
RDF in the MD simulation: this means accuracy of about 3 to 4
digits is already achieved in the QHNC-MD calculation of Run-1.  In
a consistent way to the convergence of the RDF, a good convergence
of the effective ion-ion potential $v_{\rm eff}(r)$ is achieved
in Run-1.
Similar to the case of liquid Li, a good
convergence of $g_{\scriptscriptstyle\rm II}(r)$ and $v_{\rm eff}(r)$ 
is also achieved for other liquid metals.
It should be noted that the preparation phase of the QHNC-MD
method largely enhances the convergence of $v_{\rm eff}(r)$.

Concerning the treatment of the raw MD data for the ion-ion RDF, it
should be emphasized that the extension procedure\cite{BR} to obtain the RDF
for the full potential applied to the raw MD data is indispensable
for the present calculation.  This situation is illustrated in 
Fig.~\ref{fig:Ligrdiff}, where the truncation error in 
$g_{\scriptscriptstyle\rm II}(r)$ due to
the use of the cutoff potential in the MD simulation is so large
that the convergence of the QHNC-MD method will not be attained
with the raw RDF data.  In addition, the extended RDF for the full
potential is almost identical with the result of VMHNC equation for
the same potential (Figure \ref{fig:Ligrdiff}) for $r\agt 5a$,
i.e., after the third peak of RDF.  This suggests that the long
ranged Friedel oscillation of $v_{\rm eff}(r)$ typically seen for
liquid metals is essential for the detailed structure of the RDF at
long distances, and it is necessary to include the information of
the long-range part of $v_{\rm eff}(r)$ into 
$g_{\scriptscriptstyle\rm II}(r)$ in
order to obtain a self-consistent solution of the effective
ion-ion potential by the QHNC-MD calculation.

It is concluded that the convergence of the present QHNC-MD
simulation is well attained from both numerical and computational
points of view, helped by a good initial 
estimation from the VMHNC equation in the
preparation phase; the extension procedure to obtain the full-range RDF
for the uncut effective potential is also necessary to get the convergence.

\subsection{Ion-ion and electron-ion radial distribution functions}
Following the procedure mentioned above, we obtain the effective interatomic 
interaction, the ion-ion and electron-ion RDF's, the charge distribution 
$\rho(r)$ of neutral pseudoatom, 
the bound electron distribution $n_{\rm b}(r)$ forming an ion and the 
bridge functions for liquid alkali metals in a self-consistent way from the 
atomic number as the only input. In the first place, we show 
structure factors, the Fourier transform of the ion-ion RDF's. It is importa
nt 
for a detail comparison with experiment to use the MD RDF corrected for a
full potential and extrapolated to whole range of distance 
in its Fourier transform.

Figure \ref{fig:LiSQ} exhibits the structure
factors for liquid Li calculated by the QHNC-MD simulation 
in comparison with the
experimental result\cite{ExpLi}.
It is clearly seen that the present
result is in excellent agreement with the experiment,
improving the detailed structure of $S_{\rm II}(Q)$ near its second
peak compared with the result of the VMHNC equation (the final
result of the preparation phase).  This improvement on $S_{\rm
II}(Q)$ by the refinement phase essentially relies on the detailed
oscillatory behavior of the bridge function extracted from the raw
RDF of the MD simulation.  The discrepancy in the bridge function
between the MD simulation and VMHNC equation gives no serious effects 
on the first peak of the RDF.  But details of the RDF for $2a\alt r\alt 5a$ 
are rather sensitive to the oscillation of $B(r)$ in a similar way as
discussed in \cite{BR}.  Therefore the use of the extracted bridge
function $B_{\scriptscriptstyle\rm MD}(r)$ to determine the corrected RDF is
important in order to guarantee the correct behavior of the
structure factor by the Fourier transform.
The structure factors calculated for Na and K are shown in Fig.~5 
in comparison with the neutron and X-ray experiments\cite{ExpNaKCs,ExpNaK}. 
The QHNC-MD structure factor of Na is in excellent agreement with the X-ray 
result (full circles) showing a small deviation 
from the neutron data (circles) at the first peak, 
while the curves of the QHNC-MD structure factor for K lies 
between the neutron (circles) and X-ray (full circles) experimental results. 
Also, the structure factors from the QHNC-MD method for Rb and Cs are 
compared with the neutron and X-ray 
experiments\cite{ExpRbCH,ExpRb,ExpRbW} in Fig.~6. 
The first peak of Rb structure factor observed by neutron 
experiment (full circles)\cite{ExpRbCH} is shifted a little to large $Q$ 
side compared with that observed the X-ray experiment 
(circles)\cite{ExpRbW}; the QHNC-MD result 
has the same first-peak position to the X-ray data with a little different 
height and shows overall agreement with the neutron observation. 
On the other hand, the QHNC-MD structure factor of Cs becomes 
higher in the first-peak 
than the experiment\cite{ExpNaKCs} and shows a small 
deviation in the phase of oscillation 
of structure factor for the large $Q$ region compared with experiment.
In our treatment, we solve the Schr\"{o}dinger equation to 
obtain the bound-electron distribution and electron-ion RDF; 
the relativistic effect is not taken 
into account. In the case of Cs, the relativistic effect may 
contribute to the 
calculation of the structure factor, since its atomic number 
$Z_{\rm A}=55$ is rather large; there is a possibility 
that this effect may be ascribed to 
this small discrepancy between the calculated and experimental results in 
Cs structure factor. In the same way as shown in the Li structure factor, 
the QHNC-VM structure factors deviate from the QHNC-MD results near the their 
second peak for Na, K, Rb and Cs as seen Figs. 5 and 6.
This second-peak difference between the QHNC-MD and QHNC-VM structure factors 
brings about a distinct difference in the RDF's obtained from their Fourier 
transform, as can be seen from an example of Na shown in Fig.~7.
Thus, the RDF from the VMHNC equation is found not so exact as to compare 
with details of experimental results for alkali liquids; the QHNC-MD method 
is shown to produce reliable results for all alkali liquids.

Next, we proceed to discuss the electronic structure around ion.
The electron-ion RDF's from the QHNC-MD method are shown in 
Fig.~8 for the case of Rb at temperature 313 K together with the ion-ion RDF. 
The electron-ion structure factor $S_{\rm eI}(Q)$ is represented in the 
form\cite{hyd}:
\begin{equation}\label{eq:sei2}
S_{\rm eI}(Q)= {\rho(Q) \over \sqrt{Z_{\rm I}} }S_{\rm II}(Q)\,,  
\end{equation}
in terms of the ion-ion structure factor and the charge distribution of the 
pseudoatom:
\begin{equation}\label{eq:rhoQ}
\rho(Q)\equiv {n_0^e C_{\rm eI}(Q)\chi_Q^0 \over 1-n_0^e 
C_{\rm ee}(Q)\chi_Q^0 }\,\,. \\
\end{equation}
With the help of the above equations, the pseudopotential method using 
the Ashcroft model potential can evaluate the electron-ion RDF by 
inserting it in (\ref{eq:rhoQ}) instead of $-C_{\rm eI}(r)/\beta$; 
its result using the empty core radius 1.27${\rm \AA}$ has no inner-core 
structure near the origin as is expected. 
In the QHNC-MD method, we need not introduce any pseudization 
in treating the core region. Therefore, the electron-ion 
RDF $g_{\scriptscriptstyle\rm eI}(r)$ exhibits the inner-core 
structure similar to a free-atom as is shown in 
Fig.~8 in contrast with result of the pseudopotential method (full circles). 
It is interesting to note that the electron-ion RDF has an oscillation 
with an inverse phase to that of the ion-ion RDF, 
since the electrons are pushed 
away by an ion as a whole in the core region. The charge distribution 
$\rho(r)$ of the pseudoatom in a liquid Rb has a similar structure 
to the distribution $\rho_{\rm 5s}(r)$ of the $5s$-electrons in 
the free-atom; therefore, the total electron-density distribution 
$n_{\rm b}(r)+\rho(r)$ is almost same to that of a free atom as was 
indicated by Ziman. 
Also, note that the positions of the dips in the electron-ion RDF are 
coincident with those appeared in  the $5s-$electron charge 
distribution $\rho_{\rm 5s}(r)$ in 
a free-atom; these dips in the electron-ion RDF reflect 
the inner structure of an ion in a metal.
The electron-ion RDF's of liquid alkali metals (Li, Na, K, Rb and Cs) 
are shown in Fig.~10, where the inner-core structure is omitted near 
the origin in each curve.
The dip in the electron-ion RDF becomes shallower and is shifted to the 
right side indicating the growth of the core size, as the atomic number 
increases from Na to Cs: the curve of Li shows an exception to this tendency.

When the electron-ion RDF is determined by solving the wave equation for the 
self-consistent potential $v_{\rm eI}^{\rm eff}(r)$, we obtain from 
Eq.~(\ref{eq:QHNCei}) the electron-ion DCF $C_{\rm eI}(r)$, which yields 
an effective ion-ion potential $v_{\rm eff}(r)$ of 
Eq.~(\ref{eq:veff}). Corresponding to each 
curves of $g_{\rm eI}(r)$ in Fig.~10, the effective ion-ion 
potential $v_{\rm eff}(r)$ is determined for each element 
as shown in Fig.~11. It should be 
emphasized that there is no units for scaling lengths and energies to 
effective ion-ion potentials determined by the QHNC-MD method in contrast 
with those obtained by using the Ashcroft model potential\cite{Ba92}.
Similarly, no scaling features are found in the effective ion-ion potential 
obtained by the Dagens-Rasolt-Taylor method\cite{DRT}, 
which can be thought of as an approximation to the QHNC 
formulation in the sense that the ion-ion RDF is there 
replaced by the spherical vacancy in Eq.~(\ref{eq:vefeN})\cite{DRTcal}. 
Nevertheless, it is shown that for liquid alkali metals near the melting 
point the ion-ion RDF's from the QHNC-MD method can be 
scaled almost into one curve by taking the average ion 
radius $a$ in the units of length as indicated in Fig.~12 except Cs.

As the summarization of this section, we conclude that the present 
application of the QHNC-MD method to liquid alkali metals provides 
the ionic structures in excellent agreement
with the experiments within the computational capacity
available at present, enabling to handle a relatively large system
size in the MD simulation in contrast with the usual {\it ab
initio} simulation for the electron-ion mixture. So as to 
be consistent with the ionic structure specified 
by the ion-ion RDF, the QHNC-MD method is shown to give the electron-ion 
RDF $g_{\scriptscriptstyle\rm eI}(r)$, 
the charge distribution $\rho(r)$ of a pseudoatom and the density 
distribution $n_{\rm b}(r)$ of the bound electrons forming an ion in 
a liquid metal at the same time; alternatively 
this electronic structure produces the 
ion-ion potential $v_{\rm eff}(r)$ used for the MD simulation.

\section{DISCUSSION}
\label{sec:disc}
We have shown that the QHNC-MD formulation provides a very precise
description of simple liquid metals at any state 
from the atomic number $Z_{\rm A}$ as the only input: this method is 
proven to yield a first-principles calculation. Our previous 
calculations with the VMHNC equation, which has been used for input 
to the QHNC-MD method, generated the structure factors with 
a small but systematic deviation from experiments near the second peak.  
Now our QHNC-MD method is shown to correct this deviation and 
to yield results in excellent agreement with
experiments in the whole range of wavenumber $Q$.  
Even in alkali metals, there is no scaling unit of lengths and energies 
for the effective ion-ion potentials $v_{\rm eff}(r)$ determined from the
QHNC-MD method; each potential is different from other and reflects a 
difference in the bound-electron structure of an ion in each metal. 
The situation is 
different in the case of the effective potential calculated by use of 
the Ashcroft pseudopotential, which leads to give almost single potential 
curve for all alkali liquids if scaled 
with the proper unit\cite{Ba92}. However, it is interesting to note
 that the ion-ion RDF's  can be scaled by the unit of $a$, the average 
ion-sphere radius, which enables plot the results in almost single 
curve for alkali liquids (Li, Na, K, Rb) near the melting point 
except the case 
of Cs. This fact was already found in the experimental results for 
the structure factors\cite{ExpNaKCs}.

In our QHNC-MD formulation, the exchange-correlation effect expressed by 
the LFC $G(Q)$ and the LDA $\mu_{\rm XC}(n)$ are introduced from outside the
framework of our formulation: those of the jellium model, where the ion
distribution is replaced by the positive uniform background. The
jellium model gives a good description for the electrons in alkali
metals; the present success of the QHNC-MD method depends on this fact. In
addition, the structure factors of alkali metals calculated by this
method are almost independent of what kind of LFC to choose.
However, it should be noted that the LFC in the QHNC formulation
should depend on the ion configuration precisely, since it is
defined for the electron-electron DCF in the electron-ion mixture.

To compare the MD structure factors with experiments in detail, it
is important to extrapolate the MD RDF to large distances and to correct 
errors caused by the cut effective ion-ion potential used in the 
MD simulation. Our extrapolation
method\cite{BR} is shown very efficient in dealing with the raw 
MD data for a liquid
metal with ion-ion potential accompanied by a long-range
Friedel oscillation; this extrapolation method is indispensable 
to obtain a convergent solution in the QHNC-MD method. 
In order to obtain so reliable bridge function 
for the extrapolation of the MD RDF, it is necessary to 
get a reasonable statistical accuracy for evaluation of the RDF; 
for this purpose, the MD simulation must be performed for at 
least several thousand particles taking 
about $10^{10}$ to $10^{11}$ samples 
\cite{BR}.

The Car-Parrinello MD (CP-MD) method is based on the same ground to
the QHNC-MD method:  the electron-ion mixture model for liquid metals and the
jellium model for electrons, which are treated by the 
DF theory. Also, the bare ion-ion interaction is taken
as a pure Coulombic in both treatment. However, in the CP-MD method, 
the bare electron-ion interaction is approximated by a 
pseudopotential, which is introduced from outside of the 
CP-MD formulation; this fact makes a contrast with our QHNC-MD 
method where it is obtained self-consistently within the framework 
of the QHNC formulation and no pseudization is necessary 
in treating the core region. As a consequence, the electron-ion RDF 
extracted from the CP-MD result does not have an inner core structure.
It should be emphasized that 
in the QHNC formulation the inner electronic structure around 
a fixed nucleus, such as 
the electron-ion RDF in the core region and the bound-electron 
distribution $n_{\rm b}(r)$ of an ion in a liquid metal, is 
determined to be 
consistent with the outer structure, that is, with the surrounding 
ion and electron configurations.
Therefore, the QHNC-MD method can be applied to a high density plasma, 
where a usual pseudopotential can not be constructed due to the fact 
the atomic structure in a highly compressed plasma state is quite 
different from that in the vacuum.
  While in the CP-MD treatment 
the electron distribution is determined for the multi-ion 
configuration and the ions are considered to be interacting via 
many-body forces, our approach deals with the one-center problem 
to determine the electron and ion distributions around a fixed ion 
in a liquid metal.
As this connection, it should be kept in mind that the
electron-ion mixture can be exactly taken as a quasi one-component
system interacting only via a {\it pairwise} interaction in 
the evaluation of the ion-ion RDF for simple liquid metals.

The advantage of the present method against the CP-MD method based on the 
usual pseudopotential theory can be summarized as follows:  (1) The
present procedure is capable to handle a large system size ($\sim$
$10^3$ to $10^4$ particles) in the MD simulation within the
computational resources available at present, helped by the good
initial guess with the VMHNC approximation for solving the QHNC
equation.  (2) In the QHNC-MD method, the many-body forces and 
nonlinear effect in the
electron screening are taken into account automatically in the form
of a pairwise interaction in such a way that the nonlinear
pseudopotential is constructed in terms of $C_{\rm eI}(r)$.  (3) By
setting up an additional integral equation for $C_{\rm ee}(r)$, our
method can treat the case where the jellium model for the electrons
in a metal breaks down, that is, where the exchange-correlation effect
begins to depend on the the ion configuration\cite{hyd}.
Therefore, (4) our method is applicable to high density plasmas
where the ionic structure becomes significantly so different from a
free atom due to high compression that the usual pseudopotential
theory cannot be applied.

The QHNC-MD method is developed on the base of the ion-electron model, 
where the bound electrons are assumed to be clearly distinguished 
from the conduction electrons and
the ions are so rigid and so small that the ion-ion bare 
interaction is taken as a pure Coulombic 
$v_{\scriptscriptstyle\rm II}(r)=(Z_{\rm I}e)^2/r$. 
Therefore, our method is applicable only to a simple metallic system. 
In a transition metal, for example, an ``ion"  cannot be clearly 
defined since the bound electrons are not distinct from 
the conduction electron, and the overlap of ``ions" is  significant: 
many-body interactions becomes important. Our method cannot be 
applied to such a case. While the CP-MD method treat electrons in the 
multi-center problem, our method treat them as the single-center 
problem to determine the density distribution around an fixed ion 
in a liquid metal. Thus, our method cannot describe states of the 
electrons in a multi-center configuration, such as the density 
of states. Moreover, it should be noticed that an orbital of an 
ion determined by the QHNC-MD simulation may be taken as an average 
orbital in some sense compared with that determined the CP-MD 
simulation: there, many-body forces are changing at every 
time-step, while the binary ion-ion interaction remains constant 
at every time-step in the QHNC-MD simulation.  
Already, we have proposed a method to treat non-simple 
metals\cite{QHNC}; the RDF's, the ionic charge and 
the muffin-tin potential are determined by solving a single-center 
problem with a bare ion-ion interaction as an input, 
while the density of states, the bare ion-ion interaction and 
the thermodynamic properties are to be obtained as results of 
the multi-center problem with use of output from the single-center problem.

The CP-MD method can produce the electron-ion RDF and 
the electron-ion structure $S_{\rm eI}(Q)$ is obtained from 
its Fourier transform. Thus, the charge distribution $\rho(Q)$ of 
a neutral pseudoatom can be calculated from Eq.~(\ref{eq:rhoQ}) 
even in the CP-MD 
simulation, since the relation (\ref{eq:sei2}) between $\rho(Q)$ 
and $S_{\rm eI}(Q)$ is exact if a liquid metal can be treated as a 
ion-electron mixture. Also, the electron-ion DCF $C_{\rm eI}(Q)$ is 
determined from Eq.~(\ref{eq:rhoQ}) to give an effective ion-ion 
interaction $v_{\rm eff}(r)$ from Eq.~(\ref{eq:veff}); these 
quantities must be consistent 
to the ion-ion structure factor $S_{\rm II}(Q)$, if the LFC $G(Q)$ 
in the electron-electron DCF $C_{\rm ee}(r)$ is chosen exact one in 
the electron-ion mixture. In this connection, there is a point to 
notice regarding the CP-MD method that the exchange-correlation 
effect of the valence electrons is treated by the LDA approximation: 
the LFC $G(Q)$ does not appear in the CP-MD method. This consistency 
test in the CP-MD method can be exemplified by applying it to 
liquid alkali metals, which are typical examples of the 
electron-ion model. At the same time, these quantities in the CP-MD 
method can be compared with the results of the QHNC-MD method, both methods 
providing first principles calculations.

\section*{ACKNOWLEDGMENTS}
We would like to thank Center of promotion of computational 
science and engineering, Japan Atomic Energy Research Institute 
for permitting us to use a
plenty amount of computational resources on the dedicated
vector-parallel processor Monte-4.\\

\vspace{.2cm}
It is a sad duty to inform the scientific community that Dr. Shaw Kambayashi 
died on April 3, 1995 following a car accident 
just after his thirtieth birthday.

\clearpage
%
\mediumtext
\begin{table}
\caption{Parameters used in the present QHNC-MD simulations for
  liquid alkali metals.  $a=(4\pi n_0^{\rm
  I}/3)^{-1/3}$ is the ion-sphere radius; $\Gamma=\beta e^2/a$ and 
  $r_{\rm s}=(4\pi n_0^{\rm  e}/3)^{-1/3}$ are the plasma parameter 
  and the electron-sphere radius in units of the Bohr radius 
  $a_{\scriptscriptstyle\rm B}$, 
  respectively.  $R_{{\rm c}1}$ and
  $R_{{\rm c}2}$ are the cutoff length of the effective ion-ion
  potential $v_{\rm eff}(r)$ in the MD simulation for the first MD
  run and second MD run [see text], respectively.
  }
\begin{tabular}{ccccccc}
element
   & T (K) & $\Gamma$ & $r_{\rm s}$ 
   & ${\it\Delta}t$ (fs) 
   & $R_{{\rm c}1}$ ($a$) & $R_{{\rm c}2}$ ($a$)  \\
\tableline
Li & 470 & 203.1 & 3.308 & 0.940 & 5.88 & 5.88 \\
Na & 373 & 209.1 & 4.046 & 2.349 & 7.06 & 7.06 \\
K  & 338 & 185.9 & 5.024 & 3.996 & 6.76 & 6.76 \\
Rb & 313 & 187.2 & 5.388 & 6.585 & 6.81 & 6.80 \\
Cs & 303 & 180.3 & 5.781 & 8.954 & 6.83 & 6.84 \\
\end{tabular}
\label{tab:param}
\end{table}
\clearpage
%
%
\begin{figure}
\caption{Flow chart of the QHNC-MD method.  The initial
  potential $v_{\rm eff}(r)$ is determined by approximating
  $g_{\scriptscriptstyle\rm II}(r)$ in Equation 
  (\protect\ref{eq:QHNCei}) by the step
  function $\theta(r-a)$ with the ion-sphere radius $a$.}
\label{fig:flow}
\end{figure}
\begin{figure}
\caption{The consistency measure ${\it\Delta}g_i(r)$
  defined by Eq.~\protect\ref{eq:diff}) for liquid Li;
  dot-dashed curve, ${\it\Delta}g_1(r)$; dashed curve,
  ${\it\Delta}g_2(r)$.  Solid curve is the final result for the 
  ion-ion RDF $g_{\scriptscriptstyle\rm II}(r)$.}
\label{fig:gconv}
\end{figure}
\begin{figure}
\caption{Comparison of the raw MD RDF for liquid
  Li with the extended RDF: dashed curve is the raw RDF data;
  solid curve is the extended RDF with the full potential by
  Eq.~(\protect\ref{eq:QHNCii}) with $B_{\rm MD}(r)$; $\circ$ 
  is the result from the VMHNC
  equation for the present self-consistent potential.  $L$ is the
  side length of the simulation cell; $L/2=12.794a=22.40{\rm\AA}$
  in the MD simulation for liquid Li with 4,000 particles.
  The discrepancy between the raw data and extended data becomes
  clear for $r\protect\agt 4a$, i.e., after the second peak of RDF.  On the
  other hand, the extended RDF becomes identical to that of the
  VMHNC equation for $r\protect\agt 5a$, i.e., after the third peak of
  RDF.}
\label{fig:Ligrdiff}
\end{figure}
\begin{figure}
\caption{The ion-ion static structure factor $S_{\rm
  II}(Q)$ for liquid Li: solid curve, the QHNC-MD result; dotted
  curve, result of the final step of the preparation phase 
  (the QHNC-VM result);
  $\circ$, experimental result taken from \protect\cite{ExpLi}. 
  The deviation of the VMHNC result from the experiment 
  is corrected by the QHNC-MD method.}
\label{fig:LiSQ}
\end{figure}
\begin{figure}
\caption{The ion-ion static structure factors $S_{\rm
  II}(Q)$ for liquid Na (the upper) and K (the lower): solid curve, 
  the QHNC-MD result; dotted
  curve, the QHNC-VM result; $\circ$
  and $\bullet$, the neutron and X-ray experiments taken from
  \protect\cite{ExpNaKCs} and \protect\cite{ExpNaK}, respectively.}
\label{fig:NaKSQ}
\end{figure}
\begin{figure}
\caption{The ion-ion static structure factors $S_{\rm
  II}(Q)$ for liquid Rb (the upper) and Cs (the lower): solid curve, 
  the QHNC-MD result; dotted
  curve, the QHNC-MD result (the final step of the preparation 
  phase); $\circ$
  and $\bullet$, the X-ray and neutron experimental results for Rb taken from
  \protect\cite{ExpRbW} and \protect\cite{ExpRbCH}, respectively.
  Experiment result for Cs denoted by $\circ$ is taken from 
  \protect\cite{ExpNaKCs}.}
\label{fig:RbCsSQ}
\end{figure}
\begin{figure}
\caption{The ion-ion radial distribution function $g_{\scriptscriptstyle\rm
  II}(r)$ for liquid Na: solid curve, 
  the QHNC-MD result; dotted curve, the QHNC-VM result; 
  $\bullet$, experimental result taken from
  \protect\cite{Re86}. The VMHNC approximation brings 
  about small errors in the amplitude and phase 
  compared with the experimental RDF.}
\label{fig:NaRDF}
\end{figure}
\begin{figure}
\caption{The electron-ion and ion-ion radial distribution functions 
  for liquid Rb: solid curve, 
  the QHNC-MD result; dotted curves, the QHNC-VM result; 
  $\bullet$, The electron-ion RDF derived by the use of the Ashcroft 
  model potential. The electron-ion RDF has an inverse phase to the ion-ion 
  RDF in their oscillation around unity.}
\label{fig:iieiRDF}
\end{figure}
\begin{figure}
\caption{The electron charge distribution of a pseudoatom in liquid Rb: 
  solid curve, the QHNC-MD result; dashed curve, the charge distribution 
  of the 5s-electrons in a free Rb atom; $\bullet$, 
  The pseudopotential result 
  with use of the Ashcroft model potential.}
\label{fig:atom}
\end{figure}
\begin{figure}
\caption{The electron-ion RDF's for liquid alkali metals: 
  inner-core structures near the origin are omitted.}
\label{fig:eiRDFs}
\end{figure}
\begin{figure}
\caption{Effective ion-ion potentials for liquid alkali metals: 
  Scaling properties are not found in the alkali potentials.}
\label{fig:effpot}
\end{figure}
\begin{figure}
\caption{The ion-ion RDF's for liquid alkali metals: 
  The RDF's of Li, Na, K and Rb are written almost in a single curve 
  in the units of length $a$, the ion-sphere radius. The RDF of Cs 
  (the dotted curve) shows a small deviation; the dashed curve: Li. 
  The difference among the results for Na, K, Rb are indiscernible on 
  this scale.}
\label{fig:iiRDFs}
\end{figure}

\begin{references}
\bibitem{CPMD} R.\ Car and M.\ Parrinello,
  {\it Phys.\ Rev.\ Lett.}\ {\bf 55}, 2471 (1985).
\bibitem{hyd} J.\ Chihara, 
  {\it Phys.\ Rev.}\ A {\bf 33}, 2575 (1986).
\bibitem{QHNC} J.\ Chihara, 
  {\it J.\ Phys.:\ Condens.\ Matter} {\bf 3}, 8715 (1991); 
  J.\ Chihara and M.\ Ishitobi,
  {\it Molecular Simulation} {\bf 12}, 187 (1994).
\bibitem{NEmodel} J.\ Chihara,
  {\it J.\ Phys.\ C:\ Solid State Phys.}\ {\bf 18}, 3103 (1985).
\bibitem{QHLi} J.\ Chihara,
  {\it Phys.\ Rev.}\ A {\bf 40}, 4507 (1989).
\bibitem{QHNa} M.\ Ishitobi and J.\ Chihara,
  {\it J.\ Phys.:\ Condens.\ Matter} {\bf 4}, 3679 (1992).
\bibitem{QHK} M.\ Ishitobi and J.\ Chihara,
  {\it J.\ Phys.:\ Condens.\ Matter} {\bf 5}, 4315 (1993).
\bibitem{QHAl} J.\ Chihara and S.\ Kambayashi,
  {\it J.\ Phys.:\ Condens.\ Matter}, {\bf 6} 10221 (1994).
\bibitem{QDCF} J.\ Chihara,
  {\it J.\ Phys.\ C:\ Solid\ State\ Phys.}, {\bf 17} 1633 (1984).
\bibitem{BR} S.\ Kambayashi and J.\ Chihara, 
  {\it Phys.\ Rev.}\ E {\bf 50}, 1317 (1994);
\bibitem{VMHNC} Y.\ Rosenfeld,
  {\it J.\ Stat.\ Phys.}\ {\bf 42}, 437 (1986).
\bibitem{ExpLi} H.\ Olbrich, H.\ Ruppersberg and S.\ Steeb,
  {\it Z.\ Naturforsch.}\ {\bf 38A}, 1328 (1983), where numerical
  data for $S_{\rm II}(Q)$ is taken from \protect\cite{ExpRb},
\bibitem{ExpNaKCs} M.J.\ Huijben and W. van der Lugt,
  {\it Acta Cryst.}\ A {\bf 35}, 431 (1979), where numerical data of
  $S_{\rm II}(Q)$ is taken from \protect\cite{ExpRb}.
\bibitem{ExpNaK} A.J.\ Greenfield, J.\ Wellendorf and N.\ Wiser, 
  {\it Phys.\ Rev.}\ A {\bf 4}, 1607 (1971).
\bibitem{ExpRbCH} J. R. D.\ Copley and S. W.\ Lovesey, 
  {\it Int. Phys. Conf. Ser.}\ {\bf 30}, 575 (1979), where numerical data of
  $S_{\rm II}(Q)$ is taken from \protect\cite{ExpRb}.
\bibitem{ExpRb} W. van der Lugt and B.P.\ Alblas,
  ``Structure factor of liquid alkali metals'',
  in {\it Handbook of Thermodynamics and Transport Properties of
  Alkali Metals}, R.W.\ Ohse, ed, Blackwell Scientific
  Publications, Oxford, 1985, pp.\ 316-317.
\bibitem{ExpRbW} Y.\ Waseda
  {\it The Structure of Non-Crystalline Materials},
  McGrow-Hill, New York, 1980.
\bibitem{Hoover} W.G.\ Hoover, A.J.C.\ Ladd, and B.\ Moran,
  {\it Phys.\ Rev.\ Lett.} {\bf 48}, 1818 (1982);
  W.G.\ Hoover, {\it Molecular Dynamics}, 
  Springer-Verlag, Berlin, 1986.
\bibitem{Algorithm} B.\ Bernu,
  {\it Physica} {\bf 122A}, 129 (1983).
\bibitem{Monte4} K.\ Asai, K.\ Higuchi, M.\ Akimoto, H.\ Matsumoto
  and Y.\ Seo,
  ``JAERI Monte Carlo Machine'',
  in {\it proceedings of the Joint International Conference on
  Mathematical Methods and Supercomputing in Nuclear
  Applications},
  H.\ K\" unster, E.\ Stein and W.\ Werner, eds,
  Kernforschungszentrum, Karlsruhe, 1993, pp.\ 341.
\bibitem{Ng} K.C.\ Ng,
  {\it J.\ Chem.\ Phys.}\ {\bf 61}, 2680 (1974).
\bibitem{GV} D.J.W.\ Geldart and S.H.\ Vosko,
  {\it Can.\ J.\ Phys.}\ {\bf 44}, 2137 (1966).
\bibitem{GL} O.\ Gunnarsson and B.I.\ Lundqvist,
  {\it Phys.\ Rev.}\ B {\bf 13}, 4274 (1976).
\bibitem{Ba92} U.\ Balucani, A.\ Torcini and R.\ Vallauri,
  {\it Phys.\ Rev.}\ A {\bf 46}, 2159 (1992).
\bibitem{DRT} L.\ Dagens, M.\ Rasolt and R.\ Taylor,
  {\it Phys.\ Rev.}\ B {\bf 11}, 2726 (1982).
\bibitem{DRTcal} S.\ Ranganathan, K.\ Pathak and Y.\ P.\ Varshni,
  {\it Phys.\ Rev.}\ E {\bf 49}, 2835 (1994).
\bibitem{Re86} L.\ Reatto, D.\ Levesque and J.\ J.\ Weis,
  {\it Phys.\ Rev.}\ A {\bf 33}, 3451 (1986).
\end{references}
\end{document}